# Doublet structures in quantum well absorption spectra due to Fano-related interference

G. Rau, P. C. Klipstein, and N. F. Johnson

*Clarendon Laboratory, Physics Department, Oxford University, Parks Road, Oxford OX1 3PU, England*

(April 16, 1998)

Abstract

In this theoretical investigation we predict an unusual interaction between a discrete state and a continuum of states, which is closely related to the case of Fano-interference. It occurs in a GaAs/Al$_x$Ga$_{1-x}$As quantum well between the lowest light-hole exciton and the continuum of the second heavy-hole exciton. Unlike the typical case for Fano-resonance, the discrete state here is *outside* the continuum; we use uniaxial stress to tune its position with respect to the onset of the continuum. State-of-the art calculations of absorption spectra show that as the discrete state approaches the continuum, a doublet structure forms which reveals anticrossing behaviour. The minimum separation energy of the anticrossing depends characteristically on the well width and is unusually large for narrow wells. This offers striking evidence for the strong underlying valence-band mixing. Moreover, it proves that previous explanations of similar doublets in experimental data, employing simple two-state models, are incomplete.

# I. INTRODUCTION

The quantum mechanical interference between a discrete state and a continuum of states is a fundamental problem in physics that was first treated by Anderson and Fano independently [1,2]. It serves as a model for a large number of very different physical systems [3]. In optical absorption spectra, this quantum interference manifests itself as a resonance with a characteristic asymmetric line shape [2], provided that both the discrete and continuous states are optically active and that the energy of the discrete state lies well within the continuum. Recently, various groups investigated such Fano-resonances both theoretically and experimentally in the optical spectra of semiconductor systems [3,4,5,6,7,8,9,10,11,12,13,14].

Here we deal with a special case of the Fano-Anderson model (FAM) that is subtly different from the classic Fano resonance case in two important ways. Firstly we consider a situation where only the discrete state is optically active, i.e. the continuum is optically *inactive*, and secondly the discrete state is *outside* the continuum but close to its onset. By considering the general FAM we show that, under certain conditions, the continuous states collectively anticross with the discrete state and consequently produce a characteristic doublet structure in the optical spectrum. One peak in the doublet derives from the discrete state outside the continuum while the other originates from a resonance within the continuum. We present a simple model which establishes the conditions required for the occurrence of this doublet structure.

The excitonic spectrum of a quantum well (QW) is a particularly suitable system for investigating effects related to Fano-Anderson-type interference, since discrete and continuous excitonic states originating from different subbands are often in close proximity in QWs, or may even be degenerate. In this work we use a GaAs/Al$_x$Ga$_{1-x}$As QW and focus specifically on the ground state of the first confined light-hole exciton, LH1-CB1, which couples very strongly to the continuum of the second confined heavy-hole exciton, HH2-CB1. This coupling has its origin purely in the mixing of the valence-bands and *not* in the 3D character of the Coulomb interaction; the latter is responsible for Fano-resonances such as the asymmetric line shape of HH3-CB1, which was reported in Refs. 10 and 12. We present a detailed theoretical analysis of this system by calculating realistic absorption spectra, taking the valence-band mixing fully into account. To shift the LH exciton close to the HH continuum we apply uniaxial stress. As it approaches the onset of the continuum, a doublet structure appears in the optical absorption spectrum with the two peaks showing anticrossing behaviour. By simplifying the Hamiltonian of our full calculations, we then demonstrate that this phenomenon can indeed be explained unambiguously on the basis of the FAM. Furthermore, we find that the minimum separation energy at the anticrossing varies characteristically with the QW width and can be up to 15 meV in narrow wells. This is unusually large and hence provides a new type of signature for the strong valence-band mixing that gives rise to the highly non-parabolic subband structure of these systems.

## II. ORIGIN OF DOUBLET

In the FAM it is assumed that the continuous states $|k\rangle$ are nondegenerate and do not couple with each other, but only with the discrete state $|x\rangle$. The eigenstates of the Fano-Anderson Hamiltonian *H* are in general a linear combination of $|x\rangle$ and $|k\rangle$, i.e.

$$|\Psi\rangle = a|x\rangle + \sum_k b_k |k\rangle \tag{1}$$

where the coefficients *a* and $b_k$ are obtained by diagonalizing *H*. To obtain the eigenenergies we need to determine the poles of the retarded Green's function. Since the FAM is one of the few exactly solvable many-body models, the denominator of the Green's function can be derived in a straightforward fashion [15]. Its zeros yield an implicit equation for the eigenenergies:

$$E - E_x - \sum_k \frac{\langle x|V|k\rangle\langle k|V|x\rangle}{E - E_k} = 0. \qquad (2)$$

Here $E_x$ and $E_k$ are the unperturbed energies of $|x\rangle$ and $|k\rangle$ respectively and $V$ is the perturbation that couples the discrete state to the continuum. The sum in Eq. (2) is the real part of the self-energy. In general Eq. (2) cannot be solved analytically. Nevertheless it is possible to show with the help of the Green's function that the discrete state gets "diluted" [2,15] if it is well within the continuum, since it becomes admixed to neighboring continuous states. If furthermore the coupling is weak and $\langle x|V|k\rangle$ varies only weakly with $k$, the discrete state becomes a quasi-bound resonance whose density distribution $|a(E)|^2$ has a Lorentzian line shape.

In this paper, however, we focus on the opposite case, i.e. when the coupling is strong and varies significantly for small changes in $E_k$. For this we assume that $|k\rangle$ forms a complete set of states with a semi-infinite energy range $E_k$, where the coupling matrix element $\langle x|V|k\rangle$ steadily falls to zero as $E_k$ goes to infinity. The coupling will then be strongest near the onset of the continuum. If most of the coupling is concentrated on a small energy range around $E_{x'}$, Eq. (2) can be approximated by replacing $E_k$ with $E_{x'}$. In this case the denominator of Eq. (2) becomes a constant and can be taken in front of the sum. Due to the completeness of the $|k\rangle$ states Eq. (2) simplifies to

$$(E_x - E)(E_{x'} - E) - \langle x|V^2|x\rangle \approx 0 \qquad (3)$$

This, however, is the characteristic equation of a two-state system with an off-diagonal matrix element $V' = \sqrt{\langle x|V^2|x\rangle}$. One state in this two-state system is the real state $|x\rangle$ and the other is a "virtual" state comprising the collective effect of the continuum. Considering the algebraic form of $V'$, we find that this virtual state can be interpreted as an image of the real state, since the discrete state is effectively coupled to itself via $V^2$ due to the completeness of the continuum. When $E_x$ and $E_{x'}$ in Eq. (3) are degenerate, the coupling will mix these two states generating a doublet with a separation energy of $\Delta E = 2V'$. We define $2V' = \Delta E_{max}$ in the following discussion.

In a real physical system the doublet separation $\Delta E$ is typically less than $\Delta E_{max}$. $\Delta E_{max}$ represents an upper limit, since the relevant coupling matrix elements will not in practice be localized to a small energy range of the continuum as was assumed above. In order to understand how the doublet separation $\Delta E$ depends on the distribution of the coupling among the continuum states, we solve the Fano-Anderson Hamiltonian with a model potential that allows us to vary this coupling distribution. Details of this simple model are given in the Appendix. As a measure for the spread of the coupling distribution we use the energy range of continuum states over which the sum $\sum_k |\langle x|V|k\rangle|^2$ acquires 90 % of its total value $V'^2$. From now on this will be called the "spread of coupling". We consider the situation where the unperturbed discrete state lies at the onset of the continuum. To illustrate the effect of the continuum on the unperturbed discrete state, we determine the imaginary part of the retarded Green's function. This corresponds to an absorption spectrum where the discrete state is optically active while the continuum is not. Results of these calculations are presented in Fig. 1(a) which shows the absorption spectra for different coupling distributions. Since $\Delta E_{max}$ is a natural energy scale for this system, all energies are given in units of $\Delta E_{max}$. The "normalized spread of coupling" in Fig. 1(a) accordingly refers to the spread of coupling as defined above divided by $\Delta E_{max}$. The continuum energy range extends from zero to $+\infty$. It can be clearly seen from Fig. 1(a) that a single resonance is obtained when the coupling spreads over a large energy range and accordingly is very weak. However, as the coupling-spread becomes sufficiently small the resonance splits into two peaks. When finally the coupling is condensed to the energy at the onset of the continuum, we obtain two sharp peaks separated by $\Delta E_{max}$ as expected from our discussion. Fig. 1(b) shows the energy separation of the doublet versus the spread of coupling, both in units of $\Delta E_{max}$ [16]. The important conclusion form this figure is that as long as the coupling-spread does not exceed an energy range of about $10\Delta E_{max}$, an unambiguous doublet structure will occur. The line shape of the doublet for this coupling-spread is shown in the inset of Fig. 1(b). In order to be able to observe this doublet structure in a real physical system, however, not only does the coupling-spread have to be small enough, but also $\Delta E_{max}$ needs to be larger than the experimental resolution. This will be discussed in more detail on the basis of excitonic absorption spectra of semiconductor quantum wells in the next section.

## III. DOUBLET IN QUANTUM WELL ABSORPTION SPECTRUM

We now discuss a candidate system for observing such doublet structures; in particular we will focus our attention on the LH1-CB1 exciton in the absorption spectrum of a GaAs/Al$_x$Ga$_{1-x}$As QW. Following Ref. 17 we can safely neglect the warping of the valence subbands. The resulting axial symmetry along the growth direction allows us to label the excitonic states according to the $z$-component of their orbital angular momentum. Since only excitons with circularly symmetric radial envelope functions are optically active, we concentrate on the $s$-states of the LH1-CB1 exciton, of which the 1$s$ state has by far the strongest oscillator strength. These LH1-CB1 *s*-states couple to excitonic states associated with other electron-hole subband pairs, in particular to the HH2-CB1 exciton. The strong coupling to this exciton originates from the underlying valence-band coupling, which in the single-particle picture causes the highly non-parabolic valence-subband dispersions found in GaAs/Al$_x$Ga$_{1-x}$As QWs [18]. Due to the conservation of total angular momentum only optically inactive HH2-CB1 excitons with *p*-orbital symmetry mix with LH1-CB1(*s*) excitons. These HH2-CB1(*p*) states form a complete set of states and can be jointly regarded as a continuum, since the discrete *p*-states are closely spaced in energy and merge with the continuum of unbound *p*-states. Hence, the strong coupling between LH1-CB1(1*s*) and HH2-CB1(*p*) excitons appears to fulfil the conditions required for the formation of the above discussed doublet structures. In an unperturbed GaAs/Al$_x$Ga$_{1-x}$As quantum well, however, LH1-CB1(1*s*) lies energetically well below the HH2-CB1(*p*) continuum. To tune it to the onset of the continuum, we apply tensile uniaxial stress along the growth-direction. We could achieve the same effect with compressive in-plane uniaxial stress, which is experimentally more feasible. However, in-plane uniaxial stress breaks the axial symmetry and accordingly causes extra stress-induced mixing effects. These extra effects would only introduce details which are unimportant for the principal objective of this paper, hence we choose the less usual method of tensile stress, since the principal results are unchanged. A complete theoretical analysis of the same system using in-plane uniaxial stress and a detailed comparison with experiment will be reported elsewhere [19].

First we performed state-of-the art calculations to determine realistic absorption spectra for the QW system described above. We adopted the model of Chao and Chuang presented in Ref. 20, which is a full valence-band mixing model for excitons in QWs. The exciton equation is solved in momentum space using the modified Gaussian quadrature method. In Fig. 2(a) we display a whole sequence of absorption spectra of a 10 nm GaAs/Al$_{0.3}$Ga$_{0.7}$As QW for increasing negative stress, which shifts LH1-CB1(1$s$) towards the HH2-CB1($p$) continuum [21]. We chose the onset of the continuum as the zero point of the energy scale. The striking feature in Fig. 2(a) is that a second peak appears at the onset of the HH2-CB1($p$) continuum as LH1-CB1(1$s$) approaches it. These two peaks clearly reveal anticrossing behaviour with an unusually large minimum separation energy of about 8.05 meV. This suggests that the second peak is indeed the image state of LH1-CB1(1$s$) caused by the strong Fano-Anderson-type interference along the lines of our earlier discussion. To prove this, we calculated the same absorption spectra using a strongly simplified model. This model only considers the LH1-CB1(1$s$) state and the HH2-CB1($p$) states, i.e. we ignore all other $s$-states of the LH1-CB1 exciton and the coupling to excitonic states associated with any other electron-hole subband pairs. The results are presented in Fig. 2(b). It can be seen that the simple model reproduces the anticrossing features very well yielding a minimum separation energy of 8.73 meV. What it cannot reproduce are the line shapes of the peaks, but for our current discussion they are not important. This shows that the occurrence of a doublet structure near the onset of the continuum can be explained within the framework of the Fano-Anderson model.

Doublet structures have been observed in QW structures before and were discussed in the literature [12,22,23]. A well-known example is the HH2-CB2 exciton doublet structure, which occurs in unperturbed QWs of certain well widths [22]. This doublet is also caused by the valence-band mixing between the LH1 and HH2 subbands, but here HH2-CB2(1$s$) is the discrete state and LH1-CB2($p$) is the continuum [12]. This case, therefore, is completely analogous to our case. Previously authors have explained this doublet essentially in terms of a mixing between *two discrete* states [12,23], i.e. the 1$s$ and the 2$p$ state; the role of the continuum was alluded to but not explored. Our results show that these explanations were incomplete; a careful analysis of the coupling matrix elements between the 1$s$ state and the $p$-states shows that the coupling between 1$s$ and 2$p$ contributes only about 20% to the total coupling, where the total coupling can be estimated by the sum $\sum_p |\langle 1s|V|p\rangle|^2$. This means that the 1$s$ state is in fact mainly coupled to the continuum and not to the 2$p$ state. Moreover we find that in the present case the spread of the coupling, which we defined in the previous section, is only of the order of 4. Hence the coupling is mainly localized near the onset of the continuum and according to Fig. 1(b) this is small enough to generate unambiguously a doublet structure. We believe our proposed explanation in terms of the Fano-Anderson model captures the physics of the doublet structures more adequately.

In order to understand what determines the minimum separation energy of the doublet structure during the anticrossing, we need to have a closer look at the matrix elements that couple LH1-CB1(1$s$) to the HH2-CB1($p$) continuum. A careful analysis shows that the strong coupling can be tracked down to a specific off-diagonal matrix element of the Luttinger-Kohn Hamiltonian [24], in agreement with the discussion in Ref. 25. In our case it acquires the following form:

$$\langle 1s|V|p\rangle = \frac{\hbar^2}{m}\sqrt{3}\gamma_3 \langle g_{\text{LH1}}|d/dz|g_{\text{HH2}}\rangle \langle \phi_{1s}|k_{\parallel}|\phi_p\rangle \qquad (4)$$

where $g$ is the hole $z$-direction envelope function and $\phi$ is the exciton in-plane envelope function. An estimate for the first matrix element on the right-hand side of Eq. (4) can be easily obtained by using infinite QW wave functions for $g_{\text{LH1}}$ and $g_{\text{HH2}}$ respectively. This yields $\langle g_{\text{LH1}}|d/dz|g_{\text{HH2}}\rangle \approx \frac{8}{3} 1/L$ which indicates that the occurrence of well-resolved doublet structures will strongly depend on the confinement of the system. As we discussed earlier, an upper limit for the minimum separation energy can be determined from $\Delta E_{max} = 2\sqrt{\langle 1s|V^2|1s\rangle}$. Here we are mainly interested in the scaling properties of the separation energy and so we approximate $\Delta E_{max}$ as $2|\langle 1s|V|1s\rangle|$, which together with Eq. (4) yields

$$\Delta E_{max} \approx \frac{8}{3}\sqrt{3}\gamma_3 \frac{\hbar^2}{m}\frac{1}{La_0} \qquad (5)$$

Here we have estimated $|\langle \phi_{1s}|k_\parallel|\phi_{1s}\rangle|$ to be approximately equal to $1/a_0$, with $a_0$ being the Bohr radius of the pseudo 2D exciton. Since $a_0$ does not strongly vary with the QW width L, we infer from Eq. (5) that the minimum separation energy at the anticrossing of the doublet structure depends inversely on the QW width. This is confirmed by Fig. 3 which displays the energy separation versus QW width in GaAs/Al$_{0.3}$Ga$_{0.7}$As QWs calculated within various approximations. The individual data points are the results of our extensive calculations that take the valence-band mixing fully into account. Due to the complicated line shapes of the peaks at the anticrossing we could not uniquely determine their centre of gravity. Instead we chose the point of maximum intensity as the peak position. To emphasise that this method is not unambiguous, we determined the separation energy for two different line broadenings, i.e. $\Gamma$=2 meV (boxes) and $\Gamma$=4 meV (circles). Although the results for these two broadenings by and large agree well with each other, they deviate for large L values. At L=200 Å the doublet separation is so small that it cannot be resolved anymore with a broadening of 4 meV. The solid line in Fig. 3 is the separation energy obtained from FAM type calculations, which we illustrated earlier in Fig 2(b). The dashed line, on the other hand, represents the separation energy obtained from the simplified expression of Eq. (5), except that here we determined the matrix element $\langle g_{LH1}|d/dz|g_{HH2}\rangle$ using the correct hole envelope functions for finite wells. We find excellent agreement between the results of our extensive calculations and the simple models for a broad range of QWs. For narrow QWs, however, they deviate from each other; this deviation has its origin in the proximity of HH2 to the edge of the QW. At this point the exciton becomes more 3D in character and consequently is not properly confined anymore. The overall close agreement provides further evidence that the anticrossing is indeed caused by a Fano-Anderson-type interference, since this is the only way to account for the unusually large separation energy. A two-state model, which only takes into account the discrete 1*s*- and the discrete 2*p*-state, significantly underestimates the minimum separation energy as can be seen from the dotted line in Fig. 3. In addition to the calculations for tensile uniaxial stress along the growth direction, we also performed equivalent calculations for *in-plane* uniaxial stress where we fully take the stress-induced symmetry lowering into account [19]. We found that the additional mixing of hole states only has a very small effect on the separation energies. For that reason, we include in Fig. 3 experimental results from in-plane uniaxial stress measurements for two different QWs. The downward triangle was obtained from PLE measurements on a 220 Å GaAs/Al$_{0.3}$Ga$_{0.7}$As QW presented in Ref. 26 while the upward triangle was obtained from photoreflectance measurements on a 100 Å GaAs/Al$_{0.2}$Ga$_{0.8}$As QW [19]. Experimental and theoretical results agree very well, confirming the high accuracy of our calculations.

Finally we would like to point out that the characteristic width dependence of the separation energy in addition represents a fingerprint of the Luttinger-Kohn matrix element given in Eq. (4) and hence of the underlying complex subband structure. As the LH1 subband approaches HH2 their respective energy dispersions become strongly non-parabolic; the LH1 subband acquires a very large average effective mass (due to its *electron-like* dispersion near the Brillouin zone centre) while the opposite is true for HH2. Accordingly the binding energy of an exciton attached to the LH1 subband will increase while it will decrease for the exciton attached to the HH2. Thus in this context the energy separation of the doublet can be interpreted as a consequence of differing binding energies caused by different kinetic energies. This conclusion is reinforced by the expression of Eq. (5), which represents exactly that difference in kinetic energy.

## IV. CONCLUSION

In summary, we have discussed an unusual case of the Fano-Anderson model, when the discrete state is outside a semi-infinite continuum but close to its onset. If the coupling between the discrete state and the continuum is mainly concentrated near the edge of the continuum, a doublet structure appears in the optical absorption spectrum. Hence, in this particular case, the FAM effectively behaves like a two-state system. One of the two states is a real state while the other is an "image" state.

We have been able to identify these "Fano doublets" in the excitonic absorption spectrum of a GaAs/Al$_x$Ga$_{1-x}$As QW. As the optically active LH1-CB1(1$s$) is shifted towards the optically inactive HH2-CB1($p$) continuum a doublet structure occurs in our calculated absorption spectrum, which reveals anticrossing behaviour. This can be adequately explained in terms of the FAM. Furthermore, it is possible within the framework of the FAM to derive an analytic expression that accounts for the unusually large separation energy of the anticrossing and its characteristic dependence on the width of the QW. We were also able to rule out oversimplified two-state models, which neglect the important contribution of the continuum, as an alternative explanation. The occurrence of Fano doublets in semiconductor QWs therefore represents an impressive demonstration of the importance of the valence-band mixing in coupling exciton states derived from different pairs of subbands.

**ACKNOWLEDGMENTS**

We are indebted to A. R. Glanfield for kindly allowing us to analyse his unpublished experimental data. One of us (G.R.) also would like to thank Andrei Malyshev for his valuable comments and discussions.

**APPENDIX: A SIMPLE FANO-ANDERSON MODEL**

To study the conditions that lead to the formation of the type of doublet structure discussed above, we use a simple model potential for the FAM to couple the discrete state to the continuum. The simplicity of the model allows us to vary the spread of the coupling across the continuum, which we defined earlier to be the energy range of continuous states over which the sum $\sum_k |\langle x|V|k\rangle|^2$ acquires 90 % of its total value, i.e. $\langle x|V^2|x\rangle$. For the calculations we need to discretise the continuum and hence we choose the energy spectrum of a particle in a large box, i.e. $E_k = E_1 k^2$, where $E_1 << \Delta E_{max} = 2\sqrt{\langle x|V^2|x\rangle}$. For the calculations presented in Fig. 1 the energy of the unperturbed discrete state is at the onset of the continuum, i.e. $E_x=0$. The off-diagonal matrix element that couples $|x\rangle$ to $|k\rangle$ has the following functional form:

$$\langle x|V|k\rangle = A e^{-\sqrt{\frac{E_1}{B}}k}$$

where

$$B = 0.7547 E_{spread}$$

and

$$A = \tfrac{1}{2}\sqrt{2}\Delta E_{max}\left(\frac{E_1}{B}\right)^{\frac{1}{4}}$$

Here $E_{\text{spread}}$ refers to the spread of coupling of the potential and represents the tunable parameter in this simple model. Finally, to calculate the absorption spectrum of this system, we need to determine $|a(E)|^2$ (c.f. Eq. (1)) by diagonalising the Fano-Anderson Hamiltonian.

# Figure captions

Fig. 1 : Dependence of the formation of a Fano-doublet on the spread of coupling over the continuum. Without coupling the unperturbed discrete state coincides with the onset of the continuum, which ranges from 0 to $+\infty$ in energy. All energies are scaled by $\Delta E_{max}$, the maximum splitting energy. (a) The evolution of the Fano-doublet line shape with increasing coupling-spread, (b) The splitting energy versus the spread of coupling. In the inset the line shape of the doublet is shown for a spread of coupling of $10\Delta E_{max}$.

Fig. 2 : Calculated absorption spectra of a 10 nm GaAs/Al$_{0.3}$Ga$_{0.7}$As quantum well under tensile uniaxial stress along the growth-direction using (a) a model that fully includes the valence-band mixing and (b) the Fano-Anderson model, which only takes into account the coupling between LH1-CB1(1$s$) and HH2-CB1($p$). Zero energy corresponds to the onset of the continuum.

Fig. 3 : Dependence on the GaAs/Al$_{0.3}$Ga$_{0.7}$As quantum well width of the minimum separation energy at anticrsossings of the type shown in Fig. 2 for various approximations: complete calculations, which fully take the valence-band mixing into account, for a line broadening of $\Gamma$=2 meV (boxes) and $\Gamma$=4 meV (circles), simplified calculations using the Fano-Anderson model (solid line), simple analytic solution presented in Eq. (6) (dashed line) and solution of a two-state model (dotted line), which only takes into account the coupling between LH1-CB1(1$s$) and HH2-CB1(2$p$). The upward and downward triangles represent experimental results for in-plane uniaxial stress. The downward triangle was obtained from PLE measurements on a 220 Å GaAs/Al$_{0.3}$Ga$_{0.7}$As QW presented in Ref. 21 while the upward triangle was obtained from photoreflectance measurements on a 100 Å GaAs/Al$_{0.2}$Ga$_{0.8}$As QW [21].

**Figure 1 - Rau *et al.* / Phys. Rev. B15**

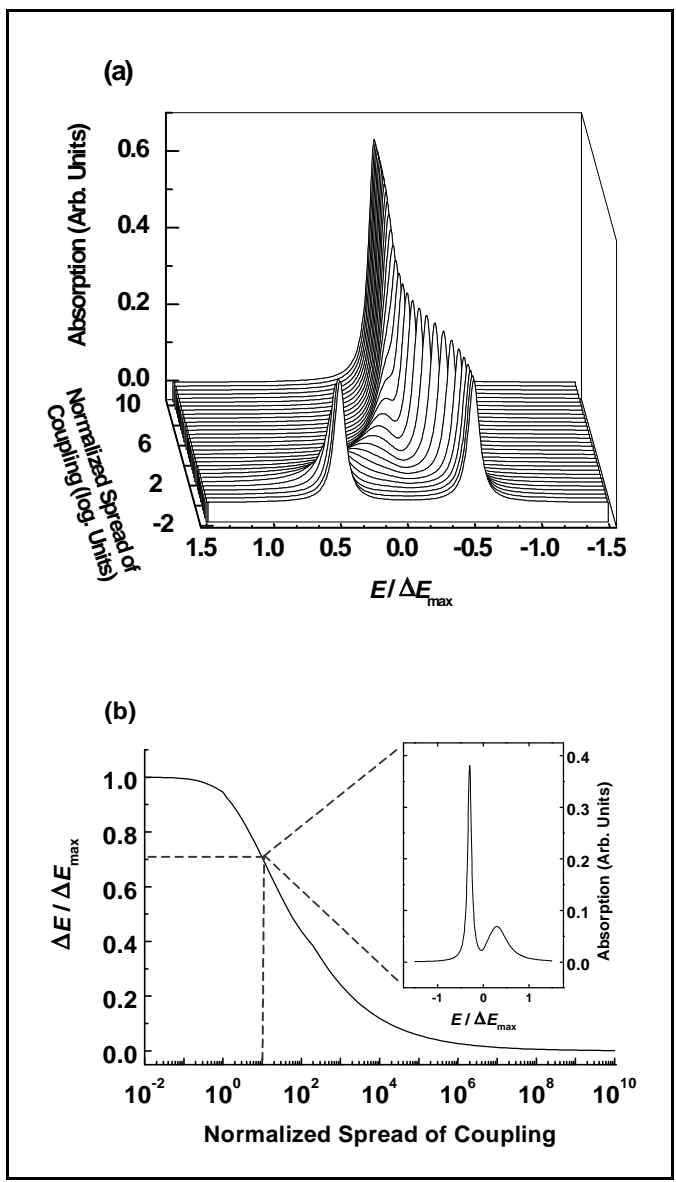

**Figure 2 - Rau *et al.* / Phys. Rev. B15**

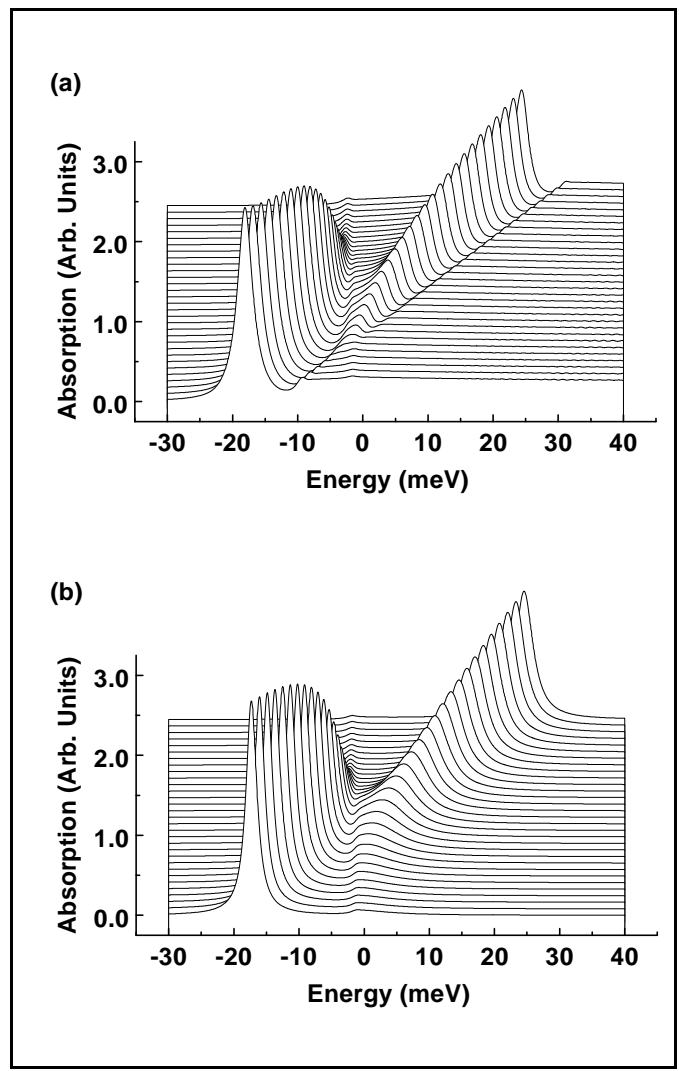

**Figure 3 - Rau *et al.* / Phys. Rev. B15**

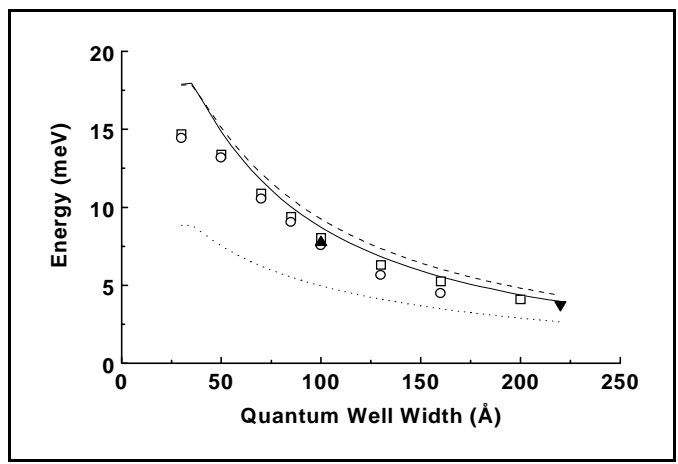